# Berry Phase Transition in Twisted Bilayer Graphene

*Johannes C. Rode\*, Dmitri Smirnov, Hennrik Schmidt, and Rolf J. Haug*

Institut für Festkörperphysik, Leibniz Universität Hannover, 30167 Hannover

ABSTRACT: The electronic dispersion of a graphene bilayer is highly dependent on rotational mismatch between layers and can be further manipulated by electrical gating. This allows for an unprecedented control over electronic properties and opens up the possibility of flexible band structure engineering. Here we present novel magnetotransport data in a twisted bilayer, crossing the energetic border between decoupled monolayers and coupled bilayer. In addition a transition in Berry phase between π and 2π is observed at intermediate magnetic fields. Analysis of Fermi velocities and gate induced charge carrier densities suggests an important role of strong layer asymmetry for the observed phenomena.

Stacked multilayer structures of graphene and other two dimensional materials have become subject of rising scientific interest over the last few years[1]. While incorporation of graphene in van der Waals heterostructures leads to exciting new phenomena[2-4], also purely graphene-based structures attracted much attention: Rich interlayer coupling phenomena like low-energy van Hove singularities and angle-dependent superlattice physics have been predicted and studied experimentally in so called twisted bilayer graphene[5-7] (TBG). TBG consists of two carbon honeycomb lattices with a certain rotational mismatch of angle $\theta$ which qualitatively divides electronic behavior in three angular ranges[7-9]: While exhibiting most complex signatures at the smallest interlayer twist $\theta \lesssim 2°$ [7-10] and effectively pristine monolayer behavior at large $\theta \gtrsim 15°$ [5,8,11], the dispersion can be understood via a perturbative model at intermediate angles[5,6,11]: In the absence of interlayer coupling the system is described by two rotationally misaligned copies of the monolayer dispersion, which displaces top and bottom layer's Dirac cones by $\Delta K = 2 \cdot \sin(\frac{\theta}{2}) \cdot K$ in reciprocal space[5,6] (with $K = \frac{4\pi}{3a}$ and $a = 246$ pm as length of graphene's lattice vector). At an interlayer hopping of magnitude $t^\theta$, the individual dispersions merge in van Hove singularities (vHs) at $\pm E_{\text{vHs}} = v_F \cdot \hbar \cdot \frac{\Delta K}{2} - t^\theta$ ($v_F$ being the Fermi velocity and $\hbar = \frac{h}{2\pi}$ the reduced Planck constant)[6,8,11-13]. Thus TBG offers the rare opportunity to study charge carriers around a divergent density of states by standard gating techniques. Additionally the energetic range between vHs features two effectively decoupled systems in closest possible vicinity, associated with phenomena like excitonic condensation, Coulomb drag or quantum capacitive screening of charge[14-19]. To date, TBG have been extensively studied by scanning tunneling microscopy resolving angle dependent moiré superstructures of wavelength

$$\lambda(\theta) = a/(2 \cdot \sin\left(\frac{\theta}{2}\right)) \tag{1}$$

and confirming the predicted vHs in spectroscopy measurements[6,8,11-13]. Another powerful tool of investigation lies in magnetotransport experiments which provide access to many of graphene's unique features[20-23]: In magnetic fields applied perpendicular to the sample plane, the Landau level spectrum for TBG is predicted to be divided into two regimes[9,24-27]: Below the vHs, assuming uniform carrier density in the two decoupled layers, Landau levels follow the energetic sequence of a single layer $E_N = \text{sgn}(N) \cdot v_F\sqrt{2e\hbar B|N|}$ but appear at doubled filling factors $\nu = \frac{n \cdot h}{B \cdot e} = N \cdot 8$ due to the additional twofold layer degeneracy (e being the elementary charge, $B$ magnetic flux density, $N$ an integer and $n$ the charge carrier density). The Fermi surface in this scenario consists of four cyclotron orbits, enclosing one Dirac point each ($K$, rotationally displaced $K_\theta$ and equivalents in opposite valley $K'$ and $K'_\theta$). This corresponds to a topological winding number of $w = \pm 1$ and a Berry phase of $\phi = \pi$ [28]. Above the vHs, different coupling models predict different scenarios: Ref. 27 finds a change in carrier polarity within the conduction (valence) band upon crossing the vHs. Cyclotron orbits now enclose a holelike (electronlike) pocket originating from the $\Gamma$-point of the superlattice mini Brillouin zone, which leads to secondary Landau fans[9,27] at a Berry phase of $\phi = 0$. In contrast, refs. 24, 25 find a continuation of the original Landau fan at modified filling factors $\nu = (N_+ + \frac{1}{2}) \cdot 4$ (with $N_+$ as nonzero integer) like in a Bernal stacked bilayer[22] ($\theta$=0°). This scenario works in the extended zone scheme and neglects commensuration effects[24]. Cyclotron orbits around $K$ and $K_\theta$ merge into one above the vHs (same for $K'$ and $K'_\theta$), now enclosing two Dirac points, which corresponds to $w = \pm 2$ and a Berry phase of $\phi = 2\pi$ [28]. The distinguishing experimental factor for one[9,27] or the other manner of coupling and quantization[24,25] might be found in the rigorosity and particular formation of the superlattice. Lattice distortions and relaxations undergo qualitative changes towards smaller angles[29] and will further depend on the choice of substrate, which may decide between the superlattice´s mini Brillouin zone and the rotated layers´ original Brillouin zones as dominant scale in k-space (see e.g. ref. 7. for the former, leading to backfolding phenomena in small angle TBG). The regime of layer decoupling has been extensively studied in experiment[17-19,30]: Most importantly, electrical gating (top or bottom gate) lifts layer degeneracy, which shows in two superposed sets of monolayerlike Shubnikov-de Haas (SdH) oscillations in longitudinal resistance[17-19,30]. The coupled regime on the other hand remains quite unexplored in comparison: Besides a recent publication[31] on higher energy bands beyond the reach of standard dynamic gating techniques, there has been one report on Bernal-bilayer-like Quantum Hall data in a TBG, which is in line with the second above described model[24,25]. We here present further evidence for the according scenario, witnessing the corresponding Berry phase transition within a primary Landau fan for the first time.

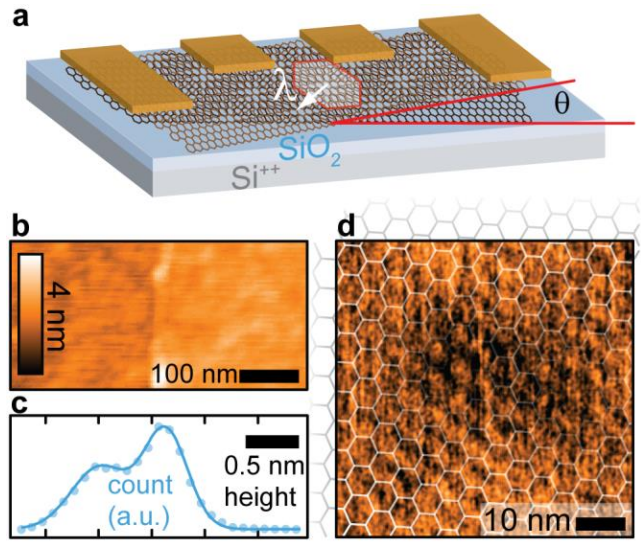

**Figure 1.** *(a) Sketch of a graphene bilayer with angle θ between top and bottom layers´ symmetry directions (red lines) and superlattice of wavelength λ (white arrow). The sample lies on a substrate of doped Si capped with $SiO_2$ and is contacted in a longitudinal setup. (b) AFM topography of the step between TBG (right) and monolayer graphene (left). (c) Dots: Histogram of pixel count over relative height for the topography image in (b). Line: Sum of two Gaussian distributions, fit to data. (d) Friction force plot of an AFM scan on the TBG with overlain honeycomb pattern as guide to the eye for the resolved superlattice.*

Graphene samples are prepared by mechanical exfoliation of natural graphite onto a substrate of SiO$_2$. Some flakes fold over during this procedure, yielding twisted layers which are processed and contacted for electrical measurements as sketched in Fig. 1a. Fig. 1b shows atomic force microscopy (AFM) topography data over the step between TBG and the uncovered monolayer, revealing a height difference of 6.2 ± 0.2 Å as evident in the histogram in Fig. 1c, fit by a double Gaussian distribution. Note that this value is larger than the interlayer spacing in graphite, which is ascribed to the different stacking arrangements[10,29,32-34]. Fig. 1d shows the torsional signal of an AFM scan on the twisted bilayer under investigation, which reveals a periodic structure of 5.7 ± 0.2 nm wavelength fit by an overlain honeycomb pattern. Using eq. 1 the corresponding twist angle $\theta$ can be calculated as 2.5 ± 0.1 °.

Fig. 2a shows an overview of longitudinal resistance vs. perpendicular magnetic field and the total charge carrier density in both layers $n_{tot}$ induced via the backgate at a temperature of 1.5 K. The data show clear deviations from the commonly expected symmetric Landau fan picture[20-22] and can be divided into three regions, displaying generally different behavior (regions **I** as well as **II** behaving qualitatively analog for both polarities of charge). To demonstrate this more clearly, Fig. 2b shows the derivative $\frac{dR}{dB}$ for positive $n_{tot}$: While the lowest depicted Landau level at the border of region **III** displays monotonous evolution in the map of $B$ vs. $n_{tot}$, higher Landau levels show an unusual discontinuity around intermediate magnetic fields, separating the data into regions **I** and **II** for $n_{\text{tot}} \gtrsim 1 \cdot 10^{16}$ m$^{-2}$. To quantify this transition, Fig. 2c shows a plot of resistance vs. inverse magnetic field at $n_{\text{tot}} = 2.97 \cdot 10^{16}$ m$^{-2}$. At high magnetic fields ($B^{-1} < 0.15$ T$^{-1}$) SdH oscillations are described by a Berry phase of $\phi = 2\pi$ indicating coupled transport[24-26]. For low magnetic fields however ($B^{-1} > 0.15$ T$^{-1}$), extrapolation of extrema to a filling factor of $\nu=0$ reveals a monolayer like quantization of $\phi = \pi$ (See Supporting Information for more examples and quantitative analysis of the Berry phase).

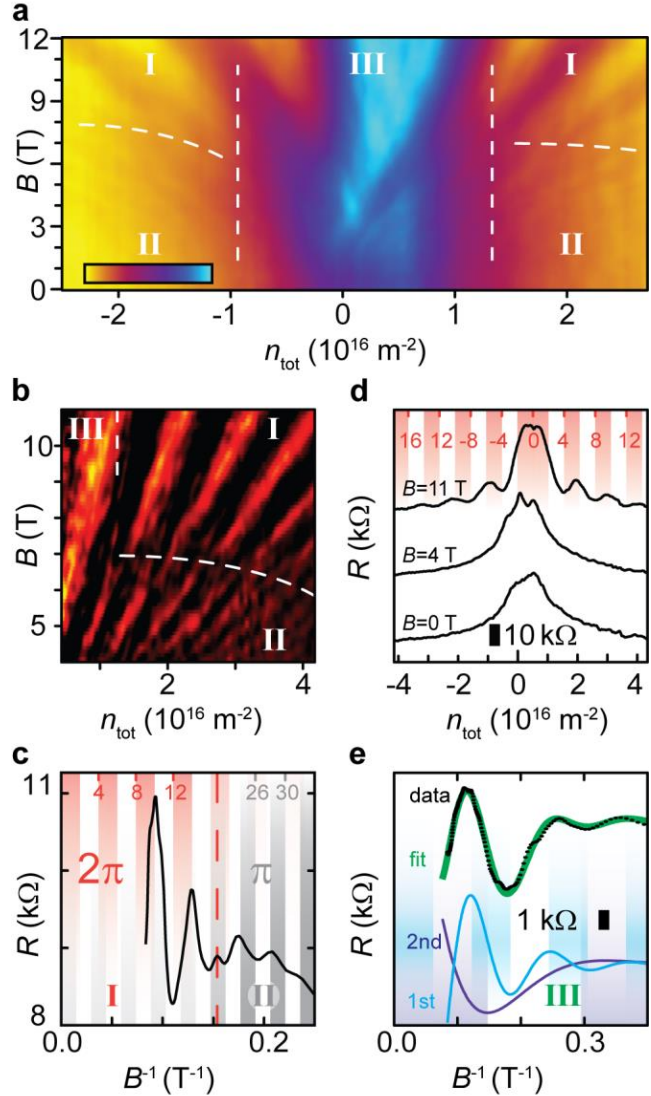

**Figure 2.** *(a) Longitudinal resistance vs. total charge carrier density and magnetic field. Dashed lines separate three regions of different magnetotransport behavior. Color scale goes from 6000 to 44000 $\Omega$ (left to right). (b) Differential longitudinal resistance $\frac{dR}{dB}$ vs. positive total charge carrier density and magnetic field. Curved horizontal line marks disruption in Landau fan between regions **I** and **II**, dashed vertical line indicates border of region **III**. (c) Resistance vs.*

*inverse magnetic field at a fixed total charge carrier density of* $2.97 \cdot 10^{16} \ m^{-2}$. *Red dashed line marks transition from* $\phi = 2\pi$ *to* $\pi$. *Colored tics at top axis indicate filling factors* $\nu$, *colored bars trace corresponding extrema in oscillations for regimes of* $2\pi$ *(red) and* $\pi$ *(gray). (d) Resistance vs. total charge carrier density at B=11 T, B=4 T and B=0 T (top to bottom, offset by 30 k$\Omega$). Red tics on top axis indicate filling factors* $\nu$ *at 11 T, red bars trace corresponding extrema in oscillations. (e) Resistance vs. inverse magnetic field at a fixed total charge carrier density of* $5.4 \cdot 10^{15} \ m^{-2}$. *Top: Black dots are data after background removal, green line is the sum of two SdH oscillations with* $\phi = \pi$, *as fitted to data. Bottom: Separately plotted contributions to the fit, colored bars indicate extrema at a monolayer-like sequence of filling factors.*

Fig. 2d shows cross sections through regions **I** and **III** at *B*=11 T and through regions **II** and **III** at *B*=4 and 0 T respectively. The resistance at *B*=11 T is modulated by pronounced SdH oscillations with $\phi = 2\pi$ confirming the high magnetic field data in Fig. 2c. At *B*=4 T oscillations in region **II** are no longer well pronounced but a double peak around $n_{tot}$=0 indicates deviation from an ordered zero mode in region **III**. The shoulder around the maximum of the field effect at *B*=0 T also indicates a more complicated behavior in the low energetic range. To explore this further, Fig. 2e shows resistance vs. inverse magnetic flux density at $n_{tot} = 5.4 \cdot 10^{15} \ m^{-2}$: A polynomial background in *B* has been removed from the data in the top half (black dots, see Supporting Information for details). The remaining oscillations are fit by the sum of two damped cosine functions (green line) which are plotted separately in the bottom half of the panel (blue, purple lines). As indicated by the colored bars, these superimposed sets of SdH oscillations exhibit a Berry phase of $\phi = \pi$, indicating parallel transport in two decoupled graphene monolayers[17-19,30].

An important theoretical prediction for the low energy dispersion between vHs is a twist angle dependent reduction in Fermi velocity following

$$\frac{v_F^{red}}{v_F^0} = 1 - 9 \cdot \left(\frac{t^\theta}{\hbar \cdot v_F^0 \cdot \Delta K}\right)^2 \qquad (2)$$

with $v_F^{red}$ and $v_F^0$ as reduced and native Fermi velocity respectively[5,8,13]. For $\theta = 2.5°$, eq. 2 yields a renormalization factor of 0.62 with the commonly found parameters $v_F^0 = 1 \cdot 10^6$ ms$^{-1}$ and $t^\theta = 0.1$ eV. Experimentally, Fermi velocities can be extracted from temperature dependence of SdH oscillations[23] as exemplified in Fig. 3a,b (See Supporting Information for examples and details of fitting procedure). Results are depicted in Fig. 3c over a range of positive total charge carrier densities, showing qualitatively different behavior for the three regions introduced in Fig. 2: Low density data points within the blue and purple areas are extracted from the two decoupled layers´ oscillations in region **III**. Both sets of velocities are clearly reduced with respect to pristine graphene. While the bottom layer data (blue, fast oscillations as exemplified in Fig. 2e) center around $0.68 \cdot 10^6$ ms$^{-1}$ close to the expected corresponding reduction value of 0.62, the top layer´s velocities (purple) lie even lower at around $0.4 \cdot 10^6$ ms$^{-1}$. As we analyze an energetic range of electrons in the bottom and holes in the top layer, we ascribe this discrepancy to electron-hole asymmetry. Like in the present case, stronger reduction in Fermi velocities on the hole side has been found in other TBG[8,13,30] and is ascribed to enhanced next-nearest-neighbor hopping[8]. As $n_{tot}$ goes across the border of region **III**, Fermi velocity starts to rise, indicating changes to the dispersion. Because region II oscillations are confined to low magnetic fields only however, further velocity data could not be reliably acquired for region **II**. High density data points in the red area stem from high magnetic field oscillations with $\phi = 2\pi$ (region **I**) and center around a constant value of $0.94 \cdot 10^6$ ms$^{-1}$ near the one of native graphene. Note that the lack of a slope in Fermi velocity over energy is indicative of massless carriers and a linear dispersion. This clearly sets our region **I** data apart from a Bernal stacked bilayer and its parabolic dispersion, commonly associated with a Berry phase of $2\pi$.

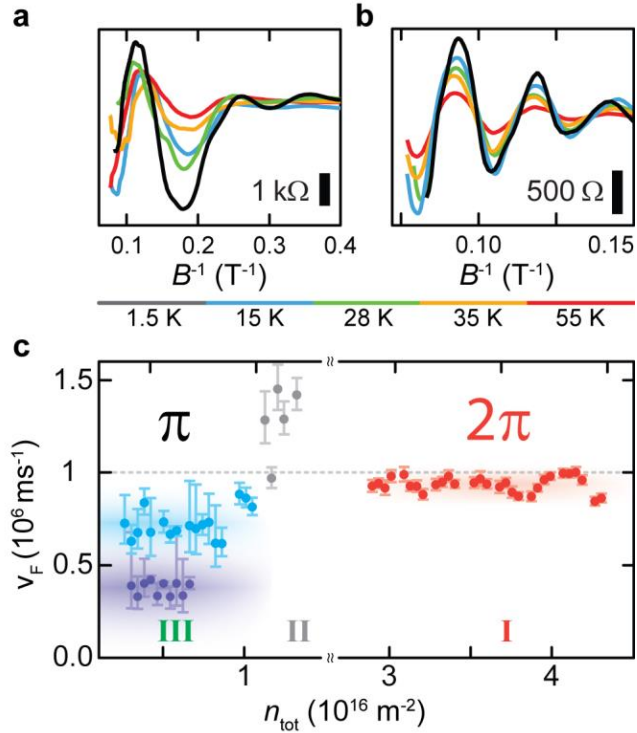

**Figure 3.** *(a,b) Resistance vs. inverse magnetic field for five different temperatures (color coded according to legend below) at an exemplary total charge carrier density of $5.4 \cdot 10^{15}\ m^{-2}$ (panel a, region **III**) and $3.99 \cdot 10^{16}\ m^{-2}$ (panel b, region **I**). The oscillations have been leveled by removal of a background resistance. (c) Fermi velocities extracted from fits to temperature dependent SdH oscillations as depicted in (a,b). Data points within the blue and purple areas in region **III** are extracted from decoupled bottom and top layers´ oscillations respectively. Rising values above $n_{tot} \approx 1 \cdot 10^{16}\ m^{-2}$ (gray dots) coincide with the transition to region **II**. Data points within the red colored area stem from the high magnetic field data in region **I**. Error bars originate from fitting uncertainty. The dashed grey line indicates the Fermi velocity of pristine graphene.*

In the range of effective decoupling (observed in region **III**), a difference $\Delta n_{tb}$ in the individual layers´ doping charge as well as application of a backgate voltage result in energetic displacement $\Delta E$ of the two layers´ Dirac cones[5,35-37]. This asymmetry in energy leads to a shift in intersection of Dirac cones in k-space by d$K$ as depicted in the schematic in Fig. 4a, leading to effective new values $\Delta K_{1,2} = \Delta K \pm 2 \cdot dK$. The renormalizing effect of interlayer coupling $t^\theta$ on the two layers´ Fermi velocities should therefore be asymmetric and can be estimated by replacing $\Delta K$ in eq. 2 with $\Delta K_{1,(2)}$ for the positive (negative) half of the bottom layer´s Dirac cone and for the negative (positive) half of the top layer´s Dirac cone respectively.

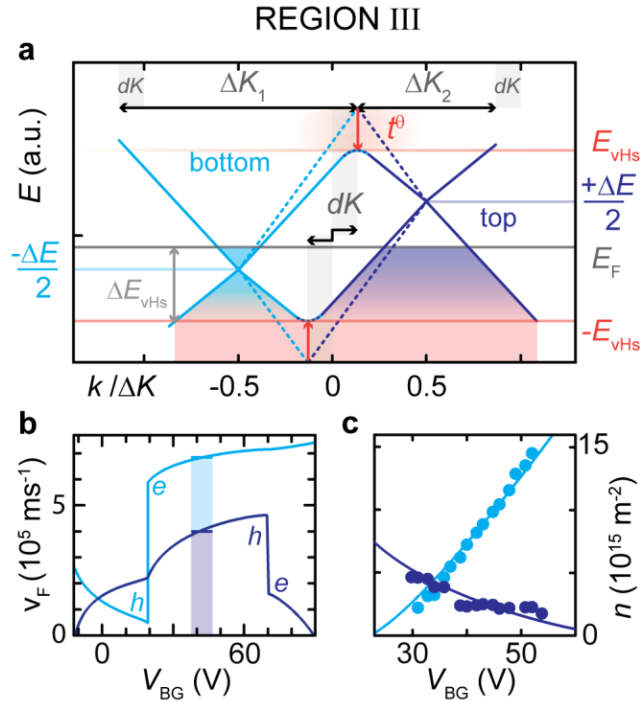

**Figure 4.** *(a) Schematic of low energy dispersion in a gated TBG. Horizontal axis cuts reciprocal plane between bottom (left) and top layer´s (right) K-points and is normalized by the magnitude of interlayer displacement vector ΔK. Dashed lines illustrate dispersion in the absence, solid lines under consideration of interlayer coupling $t^\theta$ (red arrows). The two layers´ Dirac cones intersect at k/ΔK =± dK (gray bars, black arrows) and E=±E$_{vHs}$ (red horizontal*

*lines). (b) Renormalized Fermi velocities vs. backgate voltage for bottom (blue) and top layer (purple); e, h indicate electron and hole branch respectively. Transparent bars correspond to the average value of measured Fermi velocities in bottom (blue) and top layer (purple). (c) Charge carrier densities vs. applied backgate voltage in decoupled bottom (blue) and top (purple) layers. Solid lines illustrate calculations based on screening model in main text, fit to data extracted from frequency of SdH oscillations (dots).*

This dynamic asymmetry is implemented in the established screening equations[16-19,36], which may be used to calculate top and bottom layers´ Fermi velocities, charge carrier densities and energetics in dependence on interlayer distance $d$, twist angle $\theta$, interlayer hopping energy $t^\theta$ and doping charge in the toplayer $\delta n$ (see Supporting Information). Fig. 4b shows correspondingly calculated Fermi velocities (lines) and measured values (bars) for bottom (blue) and top layer (purple) vs. applied backgate voltage $V_{BG}$. Measured charge carrier densities, extracted from frequency of SdH oscillations in both layers are depicted as dots in Fig. 4c, solid lines are calculations based on the screening model. The free parameters of doping charge and interlayer hopping have been adjusted to simultaneously fit both carrier densities $n_{b,t}$ and measured Fermi velocities $v_F^{b,t}$, yielding values of $\delta n = 1.15 \cdot 10^{16} \pm 0.10$ m$^{-2}$ and $t_e^\theta = 0.11 \pm 0.01$ eV and $t_h^\theta = 0.15 \pm 0.01$ eV for electrons and holes respectively. The interlayer hopping energy on the electron side $t_e^\theta$ is a commonly found value while its counterpart on the hole side $t_h^\theta$ lies at the topmost border of reported values[6,11-13]. The top layer´s doping $\delta n$ may be caused by deposits of processing or environmental chemicals. Extrapolation of the two layers´ densities to $V_{BG} = 0$ V yields similar values i.e. comparable degrees of doping in both layers. Such a symmetric offset in Fermi energy may also be caused by an inherent shift due to breaking of electron-hole symmetry in the TBG[5,38]. The fits to the decoupled layers´ densities are used to determine a total charge carrier density $n_{tot} = n_b + n_t$, extrapolating the TBG's capacitive coupling to the backgate away from overall charge neutrality.

In addition to the discussed modelling and data for $n_b$, $n_t$ and $n_{tot}$ in the layer-decoupled region **III**, Fig. 5 shows charge carrier concentrations extracted at higher energies. Gray dots indicate concentrations extracted from low magnetic field data at $\phi = \pi$ (region **II**), red dots in high magnetic fields at $\phi = 2\pi$ (region **I**). Solid lines are linear fits sharing an absolute slope of $6.59 \pm 0.18 \cdot 10^{14}$ m$^{-2}$V$^{-1}$ which is in good agreement with the backgate's calculative capacitive coupling constant $\alpha = 6.53 \cdot 10^{14}$ m$^{-2}$V$^{-1}$ and slope of $n_{tot}$ over $V_{BG}$.

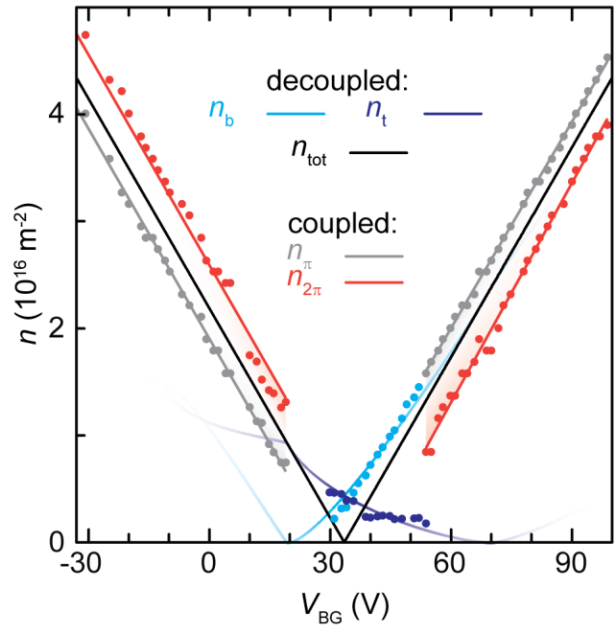

**Figure 5.** *Dots: Charge carrier concentrations extracted from frequency of SdH oscillations vs. applied backgate voltage in different regimes of quantization (see color coded legend). Blue and purple lines are fit curves based on screening model in main text, black line is correspondingly calculated total charge carrier density. Red and gray lines are linear fits to data collected in the coupled regime.*

This suggests all of the induced charge carriers filling up the examined high-energetic Landau levels, which indicates quantization of a coupled system in the corresponding ranges. Said behavior partly conforms to theory as beyond a certain energy $E_{\text{vHs}}$, layers should merge in a single system[5,24-26]. The most important prediction for this layer-coupled case is a quantization at Berry phase $\phi = 2\pi$ due to a topologically protected zero mode[24-26]. Furthermore the according charge carriers are expected to retain massless signature up to a critical magnetic flux density which would lie around 45 T for $\theta = 2.5°$ [25]. These criteria are met in region **I** featuring $\phi = 2\pi$ at a constant Fermi velocity. Although these observations comply with theory while regarded on their own, the persistence of Berry phase $\pi$ at low magnetic fields as well as deviation from $n_{\text{tot}}$ in both $n_{2\pi}$ and $n_\pi$ constitute an interesting deviation from the predicted scenario. We attribute this to strong layer asymmetry in our system, which is not accounted for in the predicted Landau quantization for TBG[24,25]. In the following we will provide a self-consistent qualitative explanation for the observed deviations from the layer-symmetric case: An important peculiarity lies in the fact, that the transition from region **III** to **II** takes place at a charge carrier concentration $n_\pi$ close to $n_\text{b}$ on the electron side (Fig. 5 at around $V_{\text{BG}}$~50V) and close to $n_\text{t}$ on the hole side (Fig. 5 at around $V_{\text{BG}}$~20V), while the opposite layer´s density is small in comparison. Note that firstly, the transition to $n_\pi$ at only the dominant layer´s density $n_\text{b}$ (or $n_\text{t}$ respectively) is consistent with the Berry phase of $\pi$ in the according oscillations, as a Berry phase of $2\pi$ would require the inclusion of both layers´ zero modes[24,25]. Secondly, exclusion of the other layer´s charge may be linked to localization due to strongly reduced Fermi velocities, when interlayer bias renders $\Delta K_{\text{eff}}$ small (compare $\Delta K_{1,2}$ in Fig. 4a) and the corresponding energy scale $E_{\text{eff}}^0$ becomes comparable to $t^\theta$ (see Fig. 6). Note that the excluded layer´s calculated Fermi velocity at the transition point (Fig. 4b) is much smaller than the dominant one´s, and even close to zero on the hole side (hole-branch of bottom layer at $V_{BG}$~20V).

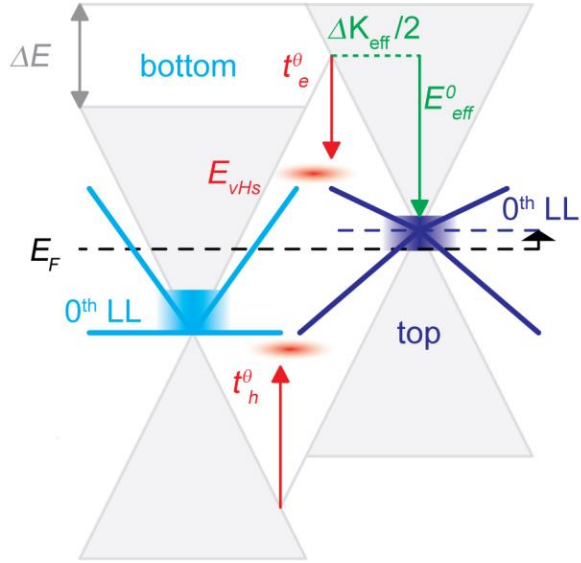

**Figure 6.** *Schematic of the electronic dispersion around the III-II-I transition at $V_{BG}=50V$ ($n_{tot} \approx 1.1 \cdot 10^{16}\ m^{-2}$) and B=6.75 T. Light gray areas are original Dirac cones, displaced by ΔE. Colored lines indicate calculated modifications to the band structure under interlayer hopping energies $t^\theta$ (red arrows). Green lines indicate momentum $\Delta K_{eff}/2$ and energy $E^0_{eff}$ from the top layer´s Dirac point to the crossing of cones. Dashed black line marks position of Fermi energy at B=0T. Purple and blue areas enclose energetic region of charge carriers contracted by the $0^{th}$ Landau level around the II-I transition at $B_{tr}=6.75$ T. Black arrow and horizontal line mark pinning of Fermi energy to the top layer´s Dirac point for $B>B_{tr}$.*

Another interesting cohesion can be found at the II-I transition. Figure 6 shows a schematic picture of the calculated dispersion at the triple point between regions I, II and III on the electron side (compare Fig. 5). The Fermi energy still lies below the vHs and, in the absence of a magnetic field, in the regime of electron conduction for the bottom and hole conduction for the top layer. Around the II-I transition at a magnetic field $B_{tr} \approx 6.75$ T (see Fig. 2) however, the zeroth Landau level of the top layer extends far enough to pin the Fermi energy (purple rectangle, Fig. 6). Thus, both layer´s zero modes may now contribute to the quantization in region I, which is in accordance with the observation of Berry phase $\phi = 2\pi$. The vanishing Fermi velocities in the top layer´s upper half cone at the transition on the electron side (top layer´s electron branch, Fig. 4) and the nearly flat dispersion in the bottom layer´s bottom half cone at the transition on the hole side (bottom layer´s hole branch, Fig. 4) are likely to be connected to the premature onset of coupling just below the calculated vHs. While the above reasoning is short of providing a closed theory on layer-asymmetric TBG, it identifies interesting cohesions in the observed phenomena, encouraging a more detailed theoretical treatment of Landau quantization in tunable TBG systems.

In summary we have studied the magnetotransport behavior in a small angle (2.5°) twisted graphene bilayer produced by folding of a single layer. The measurements show Landau quantization across the transition between a decoupled and coupled TBG system for the first time: At low energies the anticipated layer decoupling is described by a screening model. At higher energies magnetic field divides the coupled range in two regions, quantized at Berry phases of $\pi$ and $2\pi$ respectively. Together with an offset between carrier densities in the different regions we attribute this to strong asymmetry in energy and reduction of Fermi velocities between top and bottom layer.

After submission of this manuscript, very recent experimental indications[39] for the more rigorous backfolding scheme with a change of effective carrier polarity around the vHs[9,27] came to our notice. A different shaping of the superlattice due to a smaller angle as well as encapsulation of the TBG device is likely to be responsible for the manifestation of the corresponding coupling scenario[9,27] as opposed to the one evidenced in our present work[24,25].

## ASSOCIATED CONTENT

**Supporting Information.** Supplements on sample preparation and conduction of measurements; determination of Berry phase and Fermi velocities; account of screening model and implementation of dynamic asymmetry.

## AUTHOR INFORMATION

**Corresponding Author**

*E-mail: rode@nano.uni-hannover.de

**Author Contributions**

The manuscript was written through contributions of all authors. All authors have given approval to the final version of the manuscript.

**Funding Sources**

This work was supported by the DFG within the priority program SPP 1459, graphene and by the School for Contacts in Nanosystems.

## ACKNOWLEDGMENT

The authors thank Hadar Steinberg for useful discussions.

Johannes C. Rode acknowledges support from Hannover School for Nanotechnology.

REFERENCES

(1) Geim, A. K.; Grigorieva, I. V.; *Nature* **2013**, *499*, 419–425.

(2) Ponomarenko, L. A.; Gorbachev, R. V.; Yu, G. L.; Elias, D. C.; Jalil, R.; Patel, A. A.; Mishchenko, A.; Mayorov, A. S.; Woods, C. R.; Wallbank, J. R.; Mucha-Kruczynski, M.; Piot, B. A.; Potemski, M.; Grigorieva, I. V.; Novoselov, K. S.; Guinea, F.; Fal'ko, V.; Geim A. K. *Nature* **2013**, *497*, 594–597.

(3) Dean, C. R.; Wang, L.; Maher, P.; Forsythe, C.; Ghahari, F.; Gao, Y.; Katoch, J.; Ishigami, M.; Moon, P.; Koshino, M.; Taniguchi, T.; Watanabe, K.; Shepard, K. L.; Hone, J.; Kim, P. *Nature* **2013**, *497*, 598–602.

(4) Hunt, B.; Sanchez-Yamagishi, J. D.; Young, A. F.; Yankowitz, M.; LeRoy, B. J.; Watanabe, K.; Taniguchi, T.; Moon, P.; Koshino, M.; Jarillo-Herrero, P.; Ashoori, R. C. *Science* **2013**, *340*, 1427–1430.

(5) Lopes dos Santos, J. M. B.; Peres, N. M. R.; Castro Neto, A. H. *Phys. Rev. Lett.* **2007**, *99*, 256802.

(6) Li, G.; Luican, A.; Lopes dos Santos, J. M. B.; Castro Neto, A. H.; Reina, A; Kong, J.; Andrei, E. Y. *Nat. Phys.* **2010**, *6*, 109 – 113.

(7) Schmidt, H.; Rode, J. C.; Smirnov, D.; Haug, R. J. *Nat. Commun.* **2014**, *5*, 5742.

(8) Luican, A.; Li, G.; Reina, A.; Kong, J.; Nair, R. R.; Novoselov, K. S.; Geim, A. K.; Andrei, E. Y. *Phys. Rev. Lett.* **2011**, *106*, 126802.

(9) Wang, Z. F.; Liu, F.; Chou, M. Y. *Nano Lett.* **2012**, *12*, 3833−3838.

(10) Uchida, K.; Furuya, S.; Iwata, J.-I.; Oshiyama, A. *Phys. Rev. B* **2014**, *90*, 155451.

(11) Brihuega, I.; Mallet, P.; González-Herrero, H.; Trambly de Laissardière, G.; Ugeda, M. M.; Magaud, L.; Gómez-Rodríguez, J. M.; Ynduráin, F.; Veuillen, J.-Y. *Phys. Rev. Lett.* **2012**, *109*, 196802.

(12) Yan, W.; Liu, M.; Dou, R.-F.; Meng, L.; Feng, L.; Chu, Z.-D.; Zhang, Y.; Liu, Z.; Nie, J.-C.; He, L. *Phys. Rev. Lett.* **2012**, *109*, 126801.

(13) Yin, L.-J.; Qiao, J.-B.; Wang, W.-X.; Zuo, W.-J.; Yan, W.; Xu, R.; Dou, R.-F.; Nie, J.-C.; He, L. *Phys. Rev. B* **2015**, *92*, 201408(R).

(14) Min, H.; Bistritzer, R.; Su, J.-J.; MacDonald, A. H. *Phys. Rev. B* **2008**, *78*, 121401(R).

(15) Perali, A.; Neilson, D.; Hamilton, A. R. *Phys. Rev. Lett.* **2013**, *110*, 146803.

(16) S.Kim, S.; Tutuc, E. *Solid State Commun.* **2012**, *152*, 1283–1288.

(17) Schmidt, H.; Lüdtke, T.; Barthold, P.; McCann, E.; Fal'ko, V. I.; Haug, R. J. *Appl. Phys. Lett.* **2008**, *93*, 172108.

(18) Sanchez-Yamagishi, J. D.; Taychatanapat, T.; Watanabe, K.; Taniguchi, T.; Yacoby, A.; Jarillo-Herrero, P. *Phys. Rev. Lett.* **2012**, *108*, 076601.

(19) Fallahazad, B.; Hao, Y.; Lee, K.; Kim, S.; Ruoff, R. S.; Tutuc, E. *Phys. Rev. B* **2012**, *85*, 201408(R).

(20) Novoselov, K. S.; Geim, A. K.; Morozov, S. V.; Jiang, D.; Katsnelson, M. I.; Grigorieva, I. V.; Dubonos, S. V.; Firsov, A. A. *Nature* **2005**, *438*, 197-200.

(21) Zhang, Y.; Tan, Y.-W.; Stormer, H. L.; Kim, P. *Nature* **2005**, *438*, 201-204.

(22) Novoselov, K. S.; McCann, E.; Morozov1, S. V.; Fal'ko, V. I.; Katsnelson, M. I.; Zeitler, U.; Jiang, D.; Schedin, F.; Geim, A. K. *Nat. Phys.* **2006**, *2*, 177 – 180.

(23) Zou, K.; Hong, X.; Zhu, J. *Phys. Rev. B* **2011**, *84*, 085408.

(24) de Gail, R.; Goerbig, M. O.; Guinea, F.; Montambaux, G.; Castro Neto, A. H. *Phys. Rev. B* **2011**, *84*, 045436.

(25) Choi, M.-Y.; Hyun, Y.H.; Kim, Y. *Phys. Rev. B* **2011**, *84*, 195437.

(26) Lee, D. S.; Riedl, C.; Beringer, T.; Castro Neto, A. H., von Klitzing, K.; Starke, U.; Smet, J. H. *Phys. Rev. Lett.* **2011**, *107*, 216602.

(27) Moon, P.; Koshino, M. *Phys. Rev. B* **2012**, *85*, 195458.

(28) de Gail, R.; Goerbig, M. O.; Montambaux, G. *Phys. Rev. B* **2012**, *86*, 045407.

(29) van Wijk, M. M.; Schuring, A.; Katsnelson, M. I.; Fasolino, A. *2D Mater.* **2015**, *2*, 034010.

(30) Schmidt, H.; Lüdtke, T.; Barthold, P.; Haug, R. J. *Phys. Rev. B* **2010**, *81*, 121403(R).

(31) Hong, S. J.; Rodríguez-Manzo, J. A.; Kim, K. H.; Park, M.; Baek, S. J.; Kholin, D. I.; Lee, M.; Choi, E. S.; Jeong, D. H.; Bonnell, D. A.; Mele, E. J.; Drndic, M.; Johnson, A. T. C.; Park, Y. W. *Synt. Met.* **2016**, *216*, 65-71.

(32) Berashevich, J.; Chakraborty, T. *Phys. Rev. B* **2011**, *84*, 033403.

(33) Shibuta, Y.; Elliott, J. A. *Chem. Phys. Lett.* **2011**, *512*, 146-150.

(34) Jung, J.; DaSilva, A. M.; MacDonald, A. H.; Adam, S. *Nat. Commun.* **2015**, *6*, 6308.

(35) Xian, L.; Barraza-Lopez, S.; Chou, M. Y. *Phys. Rev. B* **2011**, *84*, 075425.

(36) Chung, T.-F.; He, R.; Wu, T.-L.; Chen, Y. P. *Nano Lett.* **2015**, *15*, 1203–1210.

(37) Huang, S.; Yankowitz, M.; Chattrakun, K.; Sandhu, A.; LeRoy, B. J. *arXiv:1504.08357 [cond-mat.mtrl-sci]* **2015**.

(38) Charlier, J.-C.; Michenaud, J.-P.; Lambin, P. *Phys. Rev. B* **1992**, *46*, 4540.

(39) Kim, Y.; Herlinger, P.; Moon, P.; Koshino, M.; Taniguchi, T.; Watanabe, K.; Smet, J. H. *arXiv:1605.05475 [cond-mat.mes-hall]* **2016**.

Supporting information for:

# Berry Phase Transition in Twisted Bilayer Graphene

*Johannes C. Rode, Dmitri Smirnov, Hennrik Schmidt, and Rolf J. Haug*

Institut für Festkörperphysik, Leibniz Universität Hannover, 30167 Hannover

## SAMPLE PREPARATION & MEASUREMENT SETUP

Twisted bilayer graphene (TBG) of a desired angular range can be selected based on an estimated interlayer twist derived from sample geometry[S1]. A more accurate measure of the two layer´s configuration is then obtained via resolution of the moiré pattern between the stacked lattices using an atomic force microscope.

To enable transport measurements, electrical contacts were fabricated via e-beam lithography and evaporation of Cr/Au in a longitudinal four-probe setup. A substrate of highly doped silicon, capped with 330 nm of silicon dioxide, served as a backgate to adjust the Fermi level.

Measurements were carried out in a $^4$He evaporation cryostat at temperatures down to 1.5 K in perpendicular magnetic fields up to 13 T. At a constant DC current of 500 nA, the sample resistance was measured in four-probe configuration.

## DATA ANALYSIS: BERRY PHASE

Beside the graphic approach presented in the main paper, the Berry phase $\phi$ of Shubnikov-de Haas (SdH) oscillations can be derived by a linear fit to the inverse magnetic field position of the $N$th oscillation minimum ($N$ + 0.5th maximum) over a range of $N$[S2,S3]. Fig. s1a,b shows two exemplary oscillations across regions **I** and **II** (see main paper for partition of measured range).

Extrema are determined and plotted as their index vs. position in inverse magnetic field for the two presented and three further oscillations at different charge carrier densities for region **I** and **II** respectively (Fig. s1c,d).

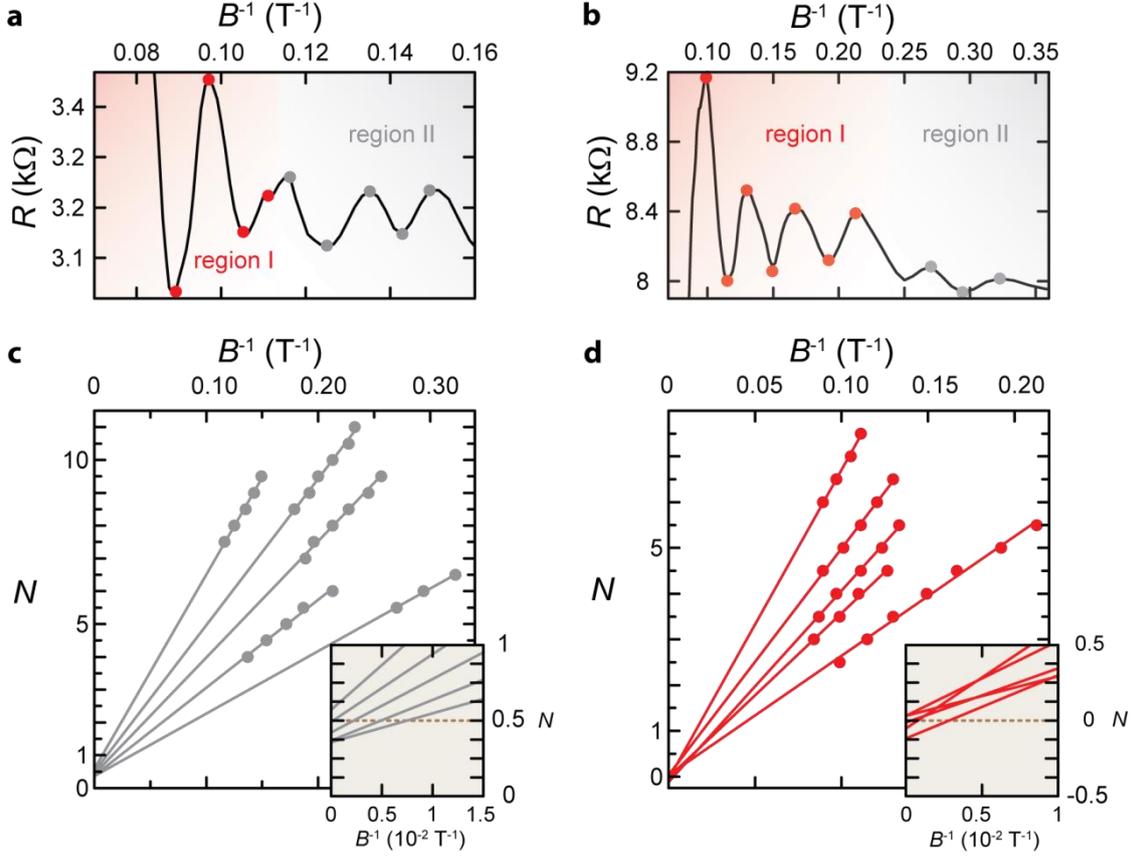

**Figure s1. (**a) Resistance vs. inverse magnetic field at $n_{tot} = -6 \cdot 10^{16}$ m$^{-2}$ and $T = 1.5$ K. Red dots mark extrema in region **I**, gray dots in region **II**. (b) Same as in (a) for $n_{tot} = -2 \cdot 10^{16}$ m$^{-2}$. (c,d) Dots: Index of extremum N vs. position in inverse magnetic field B$^{-1}$ for five oscillations at charge carrier densities between $n_{tot} = -6 \cdot 10^{16}$ (holes) and $n_{tot} = 4.25 \cdot 10^{16}$ m$^{-2}$ (electrons). The lines are linear fits to the data. Insets show close-up of the intercept region. Panel (c): region **II**, panel (d): region **I**.

As presented in the insets of fig. s1c,d the intercept of extrapolated linear fits cumulates closely around $N = 0.5$ for region **II** and $N = 0$ for region **I**. This suggests a Berry phase of $\phi = \pi$ and $\phi = 2\pi$ respectively[S2,S3], indicating a phase transition and change in topology between the analyzed regions.

# DATA ANALYSIS: FERMI VELOCITIES

**Decoupled range**

To analyze temperature damping and consecutively effective masses and Fermi velocities from the two superimposed SdH oscillations, we have used the following method: The resistance data $R$ in dependence on inverse magnetic field $B^{-1}$ has to be separated into three different contributions of the first and second oscillation and a background magnetoresistance. This is achieved by fitting the data with:

$$R(B^{-1}) = c_0 + c_1 B + c_2 B^2 + \sum_i A_i \cdot \cos(2\pi \cdot (f_i \cdot B^{-1} + 0.5) + \phi) \cdot \exp(-d_i \cdot B^{-1}) \quad (1)$$

The polynomial of second order in magnetic field $B$ with coefficients $c_{0-2}$ accounts for background magnetoresistance[S4,S5]. Two cosine functions of frequency $f_i$, damped by the exponential Dingle factor[S3,S6] with coefficient $d_i$, account for SdH oscillations with Berry phase $\phi = \pi$ in top ($i=t$) and bottom ($i=b$) layer respectively.

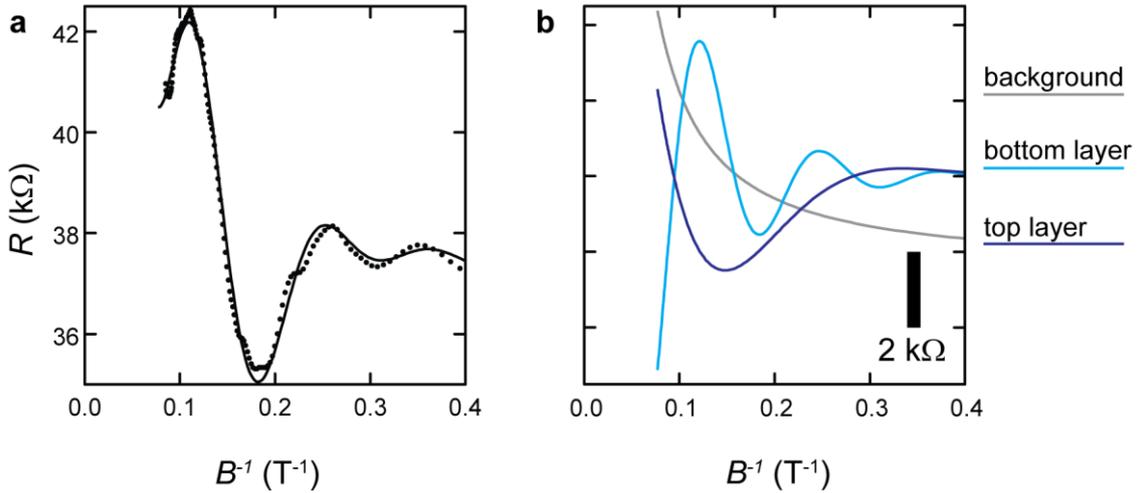

**Figure s2.** *(a) Dots: Resistance vs. inverse magnetic field at $n_{tot} = 5.4 \cdot 10^{15}$ m$^{-2}$ and $T = 1.5$ K. Line: Fit of eq. s1 to data. (b) Three additive components of the fit in (a), colored according to legend on the right.*

Fig. s2a shows a fit of eq. s1 to exemplary resistance data vs. inverse magnetic field. The fit´s three additive components are presented separately in panel b. The quality of the fit confirms the assumption of two superimposed SdH oscillations over a background. The same procedure is now applied for different temperatures at constant $n_b$, $n_t$ to enable further analysis. Fig. s3a shows data and fit curves after subtraction of the background resistances. The two constituting damped cosinusoidal components are shown in panel b. The temperature dependence of SdH resistance modulations $\Delta R_T$ should follow

$$\Delta R_T(B,T) = \frac{2\pi^2 \cdot k_B \cdot T/(\hbar \omega_c)}{\sinh(2\pi^2 \cdot k_B \cdot T/(\hbar \omega_c))} \tag{2}$$

with $k_B$ as Boltzmann constant, $\hbar$ as reduced Planck constant, and $\omega_c = \frac{e \cdot B}{m^*}$ as cyclotron frequency with effective mass $m^*$ [S2,S6].

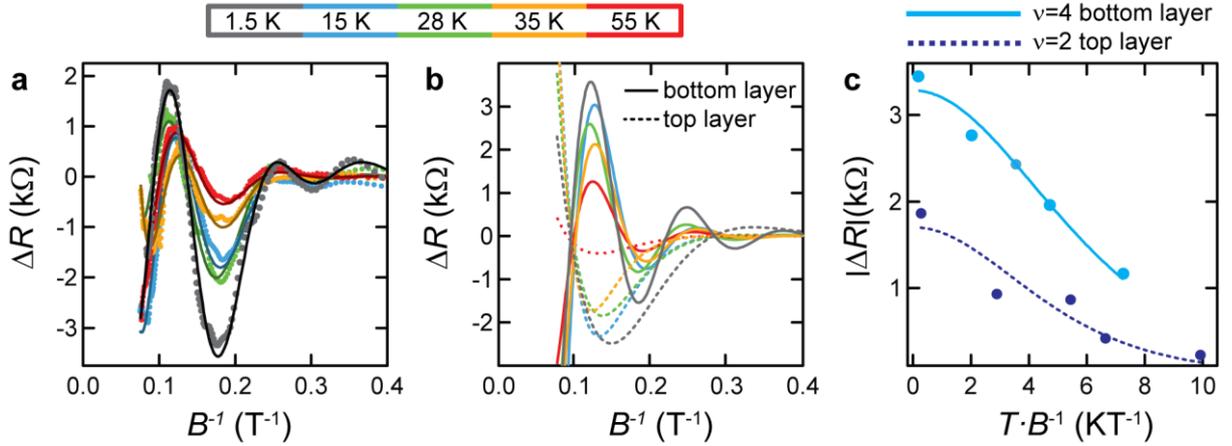

**Figure s3.** *(a) Resistance data (dots) and fit curves (lines) after background subtraction at $n_{tot} = 5.4 \cdot 10^{15}$ m$^{-2}$ for five different temperatures according to the legend on top. (b) Separate components (bottom and top layer´s contribution according to legend) of the fit curves in (a). (c) Dots: Resistance amplitude vs. $T \cdot B^{-1}$ at filling factor ν=4 (blue, bottom layer) and absolute value of resistance amplitude vs. $T \cdot B^{-1}$ at filling factor ν=2 (purple, top layer). Lines: Fits of eq. s2 to data.*

To extract effective masses for the separate layers, eq. s2 can be conveniently fit to oscillation values $\Delta R_T$ at constant filling factor $\nu = n \cdot h/(B_\nu \cdot e) = 4$ (2) for the bottom (top) layer vs. the composite variable $T \cdot B_\nu^{-1}$. According data and fits based on the oscillations in Fig. s3a,b are shown in Fig. s3c and yield effective masses of $m_b^* = 2.42 \cdot 10^{-32} \pm 1.5 \cdot 10^{-33}$ kg and $m_t^* = 2.97 \cdot 10^{-32} \pm 5.2 \cdot 10^{-33}$ kg for bottom and top layer respectively.

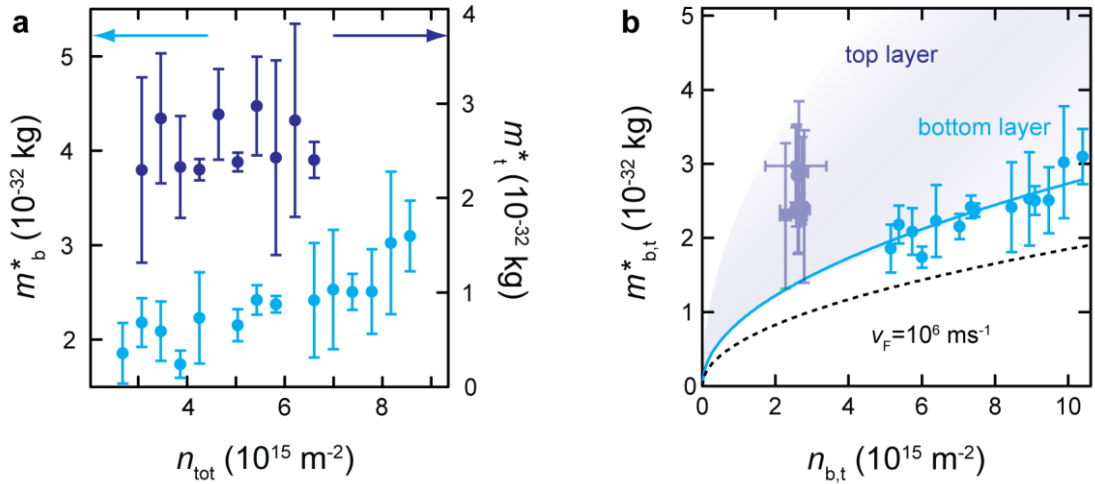

**Figure s4.** *(a) Effective mass $m^*$ vs. total charge carrier density $n_{tot}$ for bottom (blue, left axis) and top (purple, right axis) layer. Error bars stem from fitting uncertainty. (b) Dots: Effective mass $m^*$ vs. individual charge carrier densities $n_{b,t}$ for bottom (blue) and top (green) layer. Blue line marks the fit of eq. s3 to bottom layer data. Green area encloses range of possible fits to top layer data. Dashed black line marks evolution of eq. s3 for $v_F = 1 \cdot 10^6$.*

The above described procedure is repeated for a range of charge carrier densities. Thusly extracted effective masses are presented in Fig. s4. In single layer graphene the Fermi velocity $v_F$ relates to the effective mass $m^*$ as

$$v_F = \sqrt{h^2 \cdot n/(4\pi \cdot m^{*2})} \qquad (3)$$

with h as Planck constant and *n* as charge carrier density in the Dirac cones. A fit of eq. s3 to $m_b^*$ vs. $n_b$ (see Fig. s4b) yields $v_F^b = 6.84 \cdot 10^5 \pm 0.14 \cdot 10^5$ ms$^{-1}$ for the bottom layer. This is a clearly reduced value with respect to the Fermi velocity $v_F \approx 1 \cdot 10^6$ ms$^{-1}$ observed in pristine monolayer graphene, as marked by the dashed black line in Fig. s4b.

The top layer data scatter more strongly and are extracted over a small range of $n_t$ which has a flat progression in the examined region (see fig. 4c of main paper). The extracted top layer´s Fermi velocity of around $v_F^t = 4 \cdot 10^5$ ms$^{-1}$ can therefore only be seen as a rough estimate. Nevertheless it can be surely stated, that $v_F^t$ is also reduced with respect to the pristine graphene value.

**Coupled range**

To determine and remove background resistance, the data in region **I** are fit by eq. s1 with only one damped cosine term and a Berry phase of $\phi = 2\pi$. Fig. s5a shows thusly isolated oscillations for five different temperatures. A fit of eq. s2 to $\Delta R_T$ at fixed filling factor $\nu = 14$ (see example in Fig. s5b) for three different total charge carrier concentrations yields values between $m_{2\pi}^* = 3 \cdot 10^{-32}$ and $4 \cdot 10^{-32}$, presented as black squares in Fig. s5c.

As an alternative method of extraction, the simple problem of solitary oscillations in region **I** (as opposed to superposition in region **III**) allows for a global fitting procedure yielding effective masses based on temperature dependence over the whole range in $B^{-1}$ as opposed to a single filling factor. Accordingly extracted values (red circles, Fig. s5c) show good agreement with the aforementioned fixed-filling-factor results (black squares in Fig. s5c) and are used for further analysis in region **I**.

Extracted effective masses $m^*_{2\pi}$ clearly rise with charge carrier density $n_{tot}$ (Fig. s5c) which indicates a non-parabolic dispersion in the examined region **I**. A plot of $m^*_{2\pi}$ vs. corresponding charge carrier density $n_{2\pi}$ suggests a square root dependence as in eq. s3, the fit of which yields a Fermi velocity of $v_F{}^{2\pi} = 9.35 \cdot 10^5 \pm 0.08 \cdot 10^5$ ms$^{-1}$ close to the pristine monolayer case.

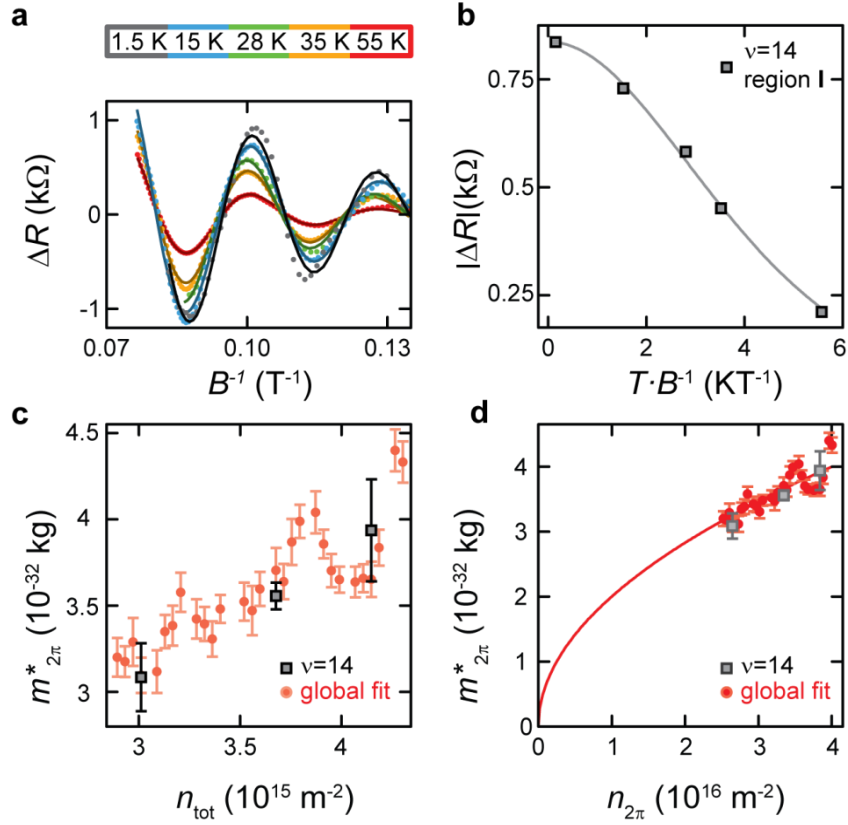

**Figure s5.** *(a) Region **I** resistance data (dots) and fit curves (lines) after background subtraction at $n_{tot} = 3.68 \cdot 10^{16}$ m$^{-2}$ for five different temperatures according to the legend on top. (b) Dots: $\Delta R_T$ at filling factor $\nu = 14$ ($B^{-1} \approx 0.1\, T^{-1}$ in panel a) vs. composite variable of temperature times inverse magnetic field. Line: Fit of eq. s2 to data. (c) Effective mass $m^*_{2\pi}$ vs. total charge carrier density $n_{tot}$. Black squares indicate extraction at $\nu = 14$, red circles via global fitting. Error bars stem from fitting uncertainty. (d) Dots: Effective mass $m^*$ vs. region **I** charge carrier density $n_{2\pi}$. The solid line marks the fit of eq. s3 to the data.*

# SCREENING MODEL

**Calculation of $n_t$**

The charge density $q_t$ induced by the backgate in the top layer gives rise to an energetic shift between top and bottom layer´s dispersion by

$$\Delta E = \frac{q_t \cdot e}{C} = \frac{d \cdot e^2}{\varepsilon_0}(n_t + \delta n) \tag{4}$$

with e as elementary charge, $C = \frac{\varepsilon_0}{d}$ as interlayer capacitance ($\varepsilon_0$ as dielectric constant, $d$ as interlayer distance) and $\delta n$ as doping charge in the top layer[S7-S9]. This way, the carrier density in the top layer $n_t$ can be calculated in dependence on variable $\Delta E$ and free parameters $d$ and $\delta n$.

**Calculation of Fermi velocities**

The position of intersection between two rotationally displaced Dirac cones (positioned at $k = \pm\frac{\Delta K}{2}$ with $k=0$ in the middle of a straight connection between Dirac points[S10]), dK is obtained by equating two linear slopes

$$-\frac{\Delta E}{2} + v_F^0 \cdot \hbar \cdot \left(k + \frac{\Delta K}{2}\right) = \frac{\Delta E}{2} - v_F^0 \cdot \hbar \cdot \left(k - \frac{\Delta K}{2}\right) \tag{5}$$

and solving to

$$k = \mathrm{d}K = \frac{\Delta E}{2 \cdot v_F^0 \cdot \hbar}. \tag{6}$$

Fermi velocities are now calculated, using $\Delta K_{1,2} = \Delta K \pm 2 \cdot \mathrm{d}K$ as described in the main text.

**Calculation of $n_b$**

The Fermi energy with respect to the top layer's Dirac point is calculated as

$$\Delta E_F^t = \text{sgn}(n_t) \cdot v_F^t \cdot \hbar \cdot \sqrt{\pi \cdot |n_t|} \tag{7}$$

with $v_F^t$ as top layer's Fermi velocity in the half-cone, crossing the Fermi level.

The Fermi energy with respect to the bottom layer's Dirac point can now be calculated as

$$\Delta E_F^b = \Delta E + \Delta E_F^t. \tag{8}$$

The bottom layer's charge carrier density follows as

$$n_b = \text{sgn}(\Delta E_F^b) \cdot \frac{(\Delta E_F^b)^2}{(v_F^b \cdot \hbar)^2 \cdot \pi} \tag{9}$$

with $v_F^b$ as bottom layer's Fermi velocity in the half-cone, crossing the Fermi level.

**Assignment of $V_{BG}$**

In the model of a parallel plate capacitor, the total charge carrier density $n_{tot}$ induced via an electrical gate with dielectric material of relative permittivity $\varepsilon_r$ and thickness $L$ couples with

$$\alpha = \frac{n_{tot}}{V_{BG}} = \frac{\varepsilon_0 \varepsilon_r}{L \cdot e} \tag{10}$$

to the gate voltage $V_{BG}$. The used wafers feature a dielectric of SiO$_2$ with $L$=330 nm and $\varepsilon_r$=3.9 which translates to a coupling constant of $\alpha = 6.53 \cdot 10^{14}$ m$^{-2}$V$^{-1}$.

The free parameter of overall charge neutrality in gate voltage, $V_{BG}^0$ is adjusted by fitting

$$n_{tot} = \alpha \cdot (V_{BG} - V_{BG}^0) \tag{11}$$

to the data.

# REFERENCES

(S1) Schmidt, H.; Rode, J. C. ; Smirnov, D.; Haug, R. J. *Nat. Commun.* **2014**, *5*, 5742.

(S2) Novoselov, K. S.; Geim, A. K.; Morozov, S. V.; Jiang, D.; Katsnelson, M. I.; Grigorieva, I. V.; Dubonos, S. V.;  Firsov, A. A. *Nature* **2005**, *438*, 197-200.

(S3) Zhang, Y.; Tan, Y.-W.; Stormer, H. L.; Kim, P. *Nature* **2005**, *438*, 201-204.

(S4) Kisslinger, F.; Ott, C.; Heide, C.; Kampert, E.; Butz, B.; Spiecker, E.; Shallcross, S.; Weber, H. B. *Nat. Phys.* **2015**, *11*, 650–653.

(S5) Zhou, Y.-B.; Han, B.-H.; Liao, Z.-M.; Wu, H.-C.; Yu, D.-P. *Appl. Phys. Lett.* **2011**, *98*, 222502.

(S6) Zou, K.; Hong, X.; Zhu, J. *Phys. Rev. B* **2011**, *84*, 085408.

(S7) Schmidt, H.; Lüdtke, T.; Barthold, P.; McCann, E.; Fal'ko, V. I.; Haug, R. J. *Appl. Phys. Lett.* **2008**, *93*, 172108.

(S8) Sanchez-Yamagishi, J. D.; Taychatanapat, T.; Watanabe, K.; Taniguchi, T.; Yacoby, A.; Jarillo-Herrero, P. *Phys. Rev. Lett.* **2012**, *108*, 076601.

(S9) Fallahazad, B.; Hao, Y.; Lee, K.; Kim, S.; Ruoff, R. S.; Tutuc, E. *Phys. Rev. B* **2012**, *85*, 201408(R).

(S10) Lopes dos Santos, J. M. B.; Peres, N. M. R.; Castro Neto, A. H. *Phys. Rev. Lett.* **2007**, *99*, 256802.